\documentclass[preprintnumbers,amsmath,amssymb,prl,aps,preprint]{revtex4-1}
\date{\today}
\usepackage{subfigure}
\usepackage{graphicx}
\usepackage{dcolumn}
\usepackage{bm}
\usepackage{float}
\usepackage{hyperref}
\usepackage{comment}
\usepackage{color}
\usepackage{lineno}

\newcommand{\kp}[1]{\textcolor{green}{#1}}
\newcommand{\kww}[1]{\textcolor{black}{#1}}

\begin{document}

\title{{Correlation induced electron-hole asymmetry in quasi-2D iridates}
}

\author{Ekaterina M. P{\"a}rschke}
\affiliation{IFW Dresden, Helmholtzstr. 20, 01069 Dresden, Germany}

\author{Krzysztof Wohlfeld}
\affiliation{Institute of Theoretical Physics, Faculty of Physics, University of Warsaw, Pasteura 5, PL-02093 Warsaw, Poland}

\author{Kateryna Foyevtsova}
 \affiliation{University of British Columbia, 6224 Agricultural Road, Vancouver, BC V6T 1Z1 Canada}
 
\author{Jeroen van den Brink}
 \affiliation{IFW Dresden, Helmholtzstr. 20, 01069 Dresden, Germany}
 \affiliation{Institute for Theoretical Physics, TU Dresden, 01069 Dresden, Germany}

\date{\today}
\begin{abstract}
We determine the motion of a charge (hole {\it or} electron) added to the Mott insulating, antiferromagnetic (AF) ground-state of quasi-2D iridates such as Ba$_2$IrO$_4$ or Sr$_2$IrO$_4$.
We show that correlation effects, calculated within the self-consistent Born approximation, render the hole and electron case very different.
An added electron forms a spin-polaron, which closely resembles the well-known cuprates, but the situation of a removed electron is far more complex.
Many-body 5$d^{4}$ configurations form which can be singlet and triplets of total angular momentum $J$ and strongly affect the hole motion between AF sublattices.
This not only has important ramifications for the interpretation of (inverse-)photoemission experiments of quasi-2D iridates but also demonstrates that the correlation physics in electron- and hole-doped iridates is fundamentally different.
\end{abstract}
\maketitle

Recently a large number of studies have been devoted to the peculiarities of the correlated physics found in the quasi-2D iridium oxides, such as e.g. Sr$_2$IrO$_4$ or Ba$_2$IrO$_4$~\cite{Kim2008, Jackeli2009, WitczakKrempa2014}. 
It was shown that this ``5$d$'' family of transition metal oxides has strong structural and electronic similarities to the famous ``3$d$'' family of copper oxides, 
the quasi-2D undoped copper oxides as exemplified by La$_2$CuO$_4$ or Sr$_2$CuO$_2$Cl$_2$~\cite{Crawford1994, Kim2008, JHKim2012}. Moreover, just as for the cuprates, the ground state of these iridates 
is also a 2D antiferromagnet (AF) and a Mott insulator~\cite{Kim2008, JHKim2012, Watanabe2014, Cao2016} - albeit formed by the $j=1/2$ spin-orbital isospins instead of the $s=1/2$ spins~\cite{Kim2008, Jackeli2009, JHKim2012}.

It is a well-known fact that the quasi-2D copper oxides turn into non-BCS superconductors when a sufficient amount of extra charge is introduced into their Mott insulating ground state~\cite{Imada1998}.
Based on the above mentioned similarities between cuprates and iridates it is natural to ask the question~\cite{WangSenthil2011} whether the quasi-2D iridates can also become superconducting upon charge doping. 
On the experimental side, very recently signatures of Fermi arcs and the ``pseudogap physics'' were found in the electron- and hole-doped iridates~\cite{Kim2014, KimNature2016, Cao2016, Yan2015} on top of the $d$-wave
gap in the electron-doped iridate~\cite{KimNature2016}. On the theoretical side, this requires studying a doped multiorbital 2D 
Hubbard model supplemented by the non-negligible spin-orbit coupling~\cite{Watanabe2010, Carter2013, Watanabe2013, Watanabe2014, Meng2014, Hampel2015, Wang2015b}. The latter is a tremendously difficult task, since even a far simpler version of this correlated model (the one-band Hubbard model) 
is not easily solvable on large, thermodynamically relevant, clusters~\cite{Maier2005}. 

Fortunately, there exists one nontrivial limit of the 2D doped Hubbard-like problems, whose solution can be obtained in a relatively exact manner. It is 
the so-called ``single-hole problem'' which relates to the motion of a single charge (hole or doublon) added to the AF and insulating 
ground state of the undoped 2D Hubbard--like model~\cite{SchmittRink1988, Kane1989}. In the case of the cuprates, such problem has been intensively studied both on the theoretical as well as the experimental (ARPES) 
side and its solution (the formation of the spin polaron) is considered a first step in understanding the motion of doped charge in the 2D Hubbard 
model~\cite{Martinez1991, Bala1995, Damascelli2003, Wang2015}. In the case of iridates several recent ARPES experiments
unveiled the shape of the iridate spectral functions~\cite{Kim2008, Wang2013, delaTorre2015, Liu2015, Nie2015, Brouet2015, Cao2016, KimNature2016, Yamasaki2016}. 
\kww{However, on the theoretical side this correlated electron problem has not been 
investigated using the above approach~\cite{Kim2008, Watanabe2014, Cao2016,BHKim2016} --
although it was suggested that the LDA+DMFT (or even LDA+U) band structure description might be sufficient~\cite{Zhang2013, Kim2008, Moser2014, delaTorre2015, Nie2015, Brouet2015}}.

Here we calculate the spectral function of the correlated strong coupling model describing the motion of a single charge doped into the AF and insulating ground state of the quasi-2D iridate, using the self-consistent Born approximation (SCBA) which is very well suited to the problem~\cite{Martinez1991, Liu1992, Sushkov1994,Brink1998, Shibata1999, Wang2015}.
The main result is that we find a fundamental difference between the motion of a single electron {\it or} hole added to the undoped iridate. Whereas the single electron added to the Ir$^{4+}$ ion locally forms a $5d^6$ configuration, adding a hole (i.e. removing an electron) to the Ir$^{4+}$ ion leads to the $5d^4$ configuration. 
\kww{(We note here that in what follows we assume that the iridium oxides are in the Mott-Hubbard regime, since the on-site Hubbard $U$ on iridium is smaller than the iridium-oxygen charge transfer 
 gap~\cite{Katukuri2012, Carter2013, Moon2009}.)}
Due to the strong on-site Coulomb repulsion, these differences in the local ionic physics
have tremendous consequences for the propagation of the doped electrons and holes. In particular: (i) in the electron case the lack of internal degrees of freedom of the added charge, forming a $5d^6$ configuration, 
makes the problem qualitatively similar to the above-discussed problem of the quasi-2D cuprates and to the formation of the spin polaron; (ii) the hopping of a hole to the nearest neighbor site does
not necessarily lead to the coupling to the magnetic excitations from $j=1/2$ AF, which is a result of the fact that the $5d^4$ configuration may have a nonzero total angular momentum
$J$~\cite{Chaloupka2016}. As discussed in the following, such a result has important consequences for our understanding of the recent and future experiments of the quasi-2D iridates. 

\begin{figure}[t!]
\includegraphics[width=0.8\columnwidth]{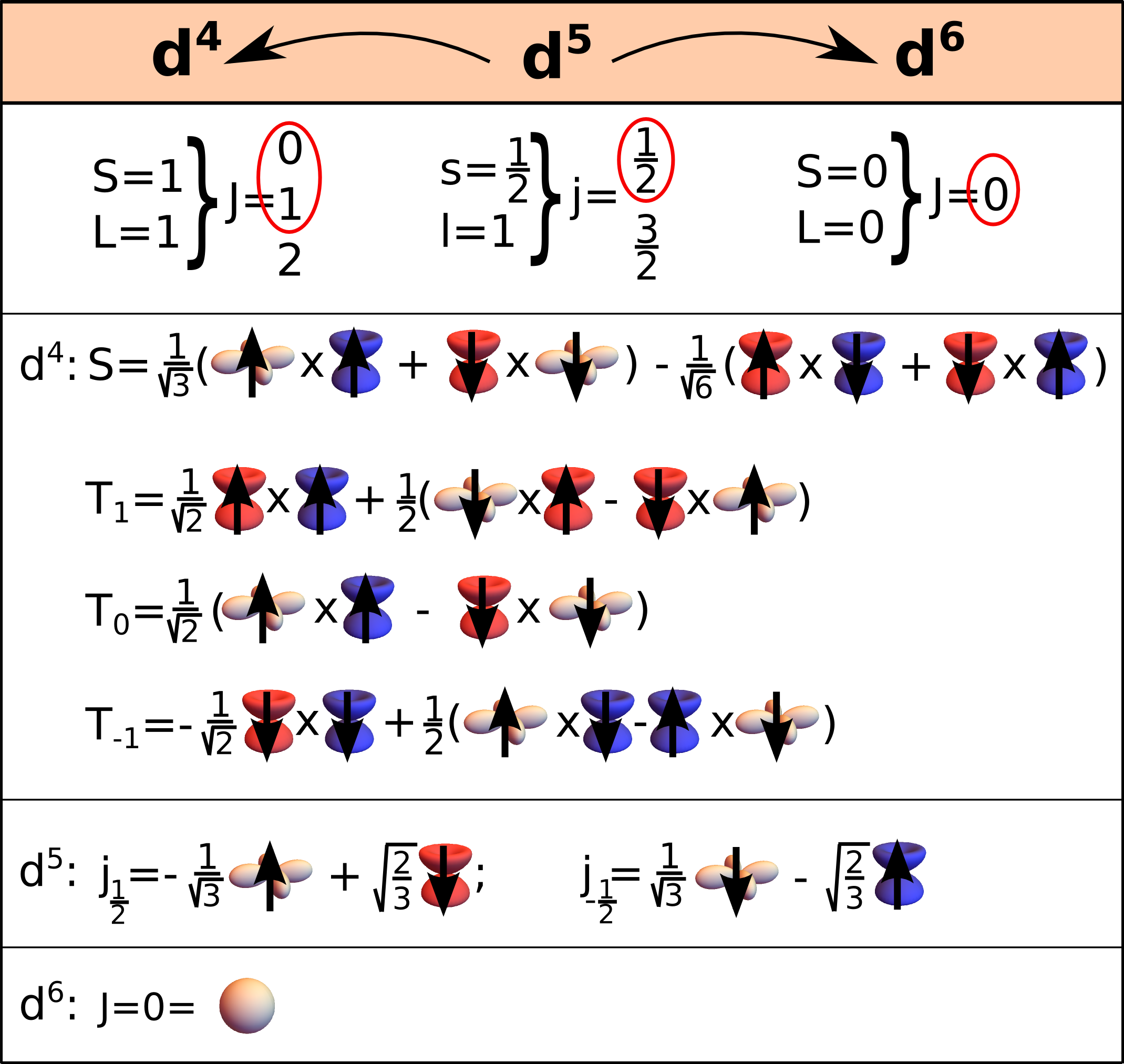}\\
\caption{
Illustration of the low energy eigenstates of $5d^4$ (relevant for the ``$5d^4$ hole'' case), $5d^5$ (relevant for the quasi-2D iridate ground state), 
and $5d^6$ (relevant for ``$5d^6$ doublon'' case) of the \kww{appropriate ionic Hamiltonian} of iridium ion. 
The red circles on top of the table indicate the states that are explicitly taken into account (see text). Blue (red) cartoon orbitals indicate 
the one-particle states with the effective angular momentum $l=1$ and $l^z=1$ $(l^z=-1)$, black arrows indicate the spin $s=1/2$ states. 
\label{fig:1}}
\end{figure}

{\bf Model} 
We begin with the low energy description of the quasi-2D iridates.
In the ionic picture \kww{(i.e. taking into account in an appropriate `ionic Hamiltonain' the cubic crystal field splitting~\cite{Sala2014}, the spin-orbit coupling~\cite{Jackeli2009}, 
and the on-site Coulomb interaction~\cite{Chaloupka2016})} the strong on-site spin-orbit coupling~\kww{$\lambda$}
splits the iridium ion $t_{2g}$ levels into the $j=1/2$ lower energy doublet (see Fig.~\ref{fig:1}) and the $j=3/2$ higher energy quartet, where
$j$ is the isospin (total angular momentum) of the only hole in the $5d^5$ iridium shell~\cite{Kim2008, Jackeli2009, WangSenthil2011, Naturecom2014Maria}.
For the bulk, the strong on-site Hubbard repulsion between holes on iridium ions needs to be taken into account which leads to the localisation of the iridium holes and the AF
interaction between their $j=1/2$ isospins in the 2D iridium plane~\cite{Jackeli2009}. 
Consequently, this Mott insulating ground state possesses 2D AF long range order with the the low energy excitations well described in the linear spin-wave approximation~\cite{Kim2012}
\begin{linenomath*}
 \begin{align}
	    \label{HamHeisenberg}
{\mathcal{H}}_{\rm mag}
=
\sum\limits_{{\bf k}} \omega_{\bf k} (\alpha^\dag_{\bf k} \alpha_{\bf k} + \beta^\dag_{\bf k} \beta_{\bf k}),
\end{align}
\end{linenomath*}
where $\omega_{\bf k}$ is the dispersion of the (iso)magnons $| \alpha_{\bf k} \rangle$ and $ | \beta_{\bf k} \rangle $  
which depends on two exchange parameters $J_1$ and $J_2$~\cite{SM}, and ${\bf k}$ is the crystal momentum.
\kww{We note here that, although the size of the experimentally observed optical gap is not large (around 500 meV~\cite{Moon2009}), it is still more than twice larger than the top of the magnon 
band in the RIXS spectra (around 200 meV)~\cite{JHKim2012, Naturecom2014Maria}. This, together with the fact that the linear spin wave theory very well describes the experimental RIXS spectra
of the quasi-2D iridates~\cite{JHKim2012, Naturecom2014Maria}, justifies 
using the strong coupling approach.}

Introducing a single electron into the quasi-2D iridates, as experimentally realised in an inverse photoemission (IPES) experiment,
leads to the creation of a single ``$5d^6$ doublon'' in the bulk, leaving the nominal $5d^5$ configuration on all other iridium sites. Since the $t_{2g}$ shell is for the $5d^6$ configuration
completely filled, the only eigenstate of the appropriate ionic Hamiltonian is the one carrying $J=0$ total angular momentum. Therefore, just as in the cuprates, the ``$5d^6$ doublon'' formed in IPES 
has no internal degrees of freedom, i.e. $ | d \rangle \equiv | J=0  \rangle$, see Fig.~\ref{fig:1}. 

Turning on the hybridization between the iridium ions leads to the hopping of the ``$5d^6$ doublon'' between iridium sites ${\bf i}$ and ${\bf j}$: 
$| 5d^5_{\bf i} 5d^6_{\bf j}\rangle \langle 5d^6_{\bf i} 5d^5_{\bf j} |$. 
It is important to realise at this point that, although such hopping is restricted to the lowest Hubbard subband of the problem,
it may change the AF configuration and excite magnons. In fact, magnons are excited during all nearest neighbor hopping processes, since
the kinetic energy conserves the total angular momentum. Altogether, we obtain the ``IPES Hamiltonian'':
\begin{linenomath*}
\begin{align}
	    \label{Hamd6}
{\mathcal{H}}_{\rm IPES}={\mathcal{H}}_{\rm mag}+{\mathcal{H}}_{t}^{\bf {\rm d}}, 
\end{align}
\end{linenomath*}
where ${\mathcal{H}}_{\rm mag}$ is defined above and the hopping of the single ``$5d^6$ doublon'' in the bulk follows from the ``spin-polaronic''~\cite{Martinez1991,Kane1989,Wohlfeld2009,Bala1995} Hamiltonian
  \begin{linenomath*}
  \begin{align}
	    \label{Hamd6parts}
	    &{\mathcal{H}}_{t}^{\bf {\rm d}} = \sum\limits_{{\bf k}}{V^{0}_{{\bf k}} \left(d^{\dagger}_{{\bf k} A}d_{{\bf k} A} + d^{\dagger}_{{\bf k} B}d_{{\bf k} B}\right)}+ \sum\limits_{{\bf k},{\bf q}} V_{{\bf k},{\bf q}}
	    \left(d^{\dagger}_{{\bf k-q} B}d_{{\bf k} A} \alpha_{\bf q}^{\dagger}+d^{\dagger}_{{\bf k-q} A}d_{{\bf k} B} \beta_{\bf q}^{\dagger} +h.c. \right),
\end{align}
\end{linenomath*}
where $A, B$ are two AF sublattices, the term $\propto V^{0}_{{\bf k}}$ describes the \kww{next nearest and third} neighbor hopping which does not excite magnons (free hopping), and 
the term $\propto V_{{\bf k},{\bf q}}$ describes the nearest neighbor coupling between the ``$5d^6$ doublon''  and the magnons as a result of the nearest neighbor electronic hopping (polaronic hopping, see above).
While the derivation and exact expressions for $V$'s are given in the Supplementary Information (SI)~\cite{SM}, we note here that they depend on the \kww{five hopping elements of the minimal 
tight binding model: $t_1$ ($t'$, $t''$) describing nearest (next-nearest, third-) neighbor hopping between the $d_{xy}$ orbitals in the $xy$ plane, $t_2$ -- the nearest neighbor
in-plane hopping between the other two active orbitals, $d_{xz}$($d_{yz}$), along the $x$($y$) direction, and $t_3$ -- the nearest neighbor hopping between
 $d_{xz}$ ($d_{yz}$) orbitals along the $y$($x$) direction.} The values of these parameters \kww{($t_1=-0.2239$ eV, $t_2=-0.373$ eV, $t'=-0.1154$ eV, $t''=-0.0595$ eV, $t_3=-0.0592$ eV)}
 are found as a best fit of this restricted tight-binding model to the LDA band structure~\cite{SM}. \kww{While in what follows we use the above set of tight-binding parameters in the polaronic
 model, we stress that the final results
are not critically sensitive to this particular choice of the model parameters.}

Next, following similar logic we derive the microscopic model for a single hole introduced into the iridate, which resembles the case encountered in the photoemission (PES) experiment.
In this case a single ``$5d^4$ hole'' is created in the bulk. Due to the strong Hund's coupling the lowest eigenstate of the appropriate ionic Hamiltonian for four $t_{2g}$ electrons has the 
total (effective) orbital momentum $L=1$ and the total spin momentum $S=1$~\cite{Khaliullin2013}. 
Moreover, in the strong spin-orbit coupled regime the $L=1$ and $S=1$ moments 
\kww{the eigenstates of such an ionic Hamiltonian are}
the lowest lying $J=0$ singlet $S$, and the higher lying $J=1$ triplets $T_\sigma$ ($\sigma=-1,0,1$\kww{, split by energy $\lambda$ from the singlet state}) and $J=2$ quintets. 
Since the high energy quintets are only marginally relevant to the low energy description in strong on-site spin-orbit coupling $\lambda$~\cite{Naturecom2014Maria} limit, 
one obtains~\cite{Chaloupka2016} that, unlike e.g. in the cuprates, the ``$5d^4$ hole'' formed in PES is effectively left with four internal degrees of freedom,
i.e. $ |{\bf h} \rangle   \equiv \{ |S \rangle, |T_{1} \rangle, |T_{0} \rangle, |T_{-1}\rangle \}$, see Fig.~\ref{fig:1}.

Once the hybridization between the iridium ions is turned on, the hopping of the ``$5d^4$ hole'' between iridium sites ${\bf i}$ and ${\bf j}$ is possible: 
$| 5d^5_{\bf i} 5d^4_{\bf j}\rangle \langle 5d^4_{\bf i} 5d^5_{\bf j} | = | 5d^5_{\bf i} \rangle \langle 5d^5_{\bf j} |  | {\bf h}_{\bf j} \rangle \langle {\bf h}_{\bf i} |$. 
Similarly to the IPES case described above, in principle such hopping  of the ``$5d^4$ hole'' may or may not couple to magnons. However, 
there is one crucial difference w.r.t. IPES:  the ``$5d^4$ hole'' can carry finite angular momentum and thus 
the ``$5d^4$ doublon''  may move between the {\it nearest neighbor sites without coupling to magnons}.
Altogether, the PES Hamiltonian reads
\begin{linenomath*}
\begin{align}
	\label{Hamd4}
{\mathcal{H}}_{\rm PES}={\mathcal{H}}_{\rm mag}+{\mathcal{H}}_{\rm SOC}+{\mathcal{H}}^{\bf {\rm \bf h}}_{t},
\end{align}
\end{linenomath*}
where ${\mathcal{H}}_{\rm SOC} =\lambda/2\sum_{{\bf k},\sigma=-1,0,1}{T^{\dagger}_{{\bf k} \sigma}T_{{\bf k} \sigma}}$
describes the on-site energy of the triplet states which follows from the on-site spin-orbit coupling $\lambda$
and the hopping of the single ``$5d^4$ hole'' in the bulk is described by the following ``spin-polaronic''~\cite{SchmittRink1988, Kane1989, Martinez1991} Hamiltonian
\begin{linenomath*}
\begin{align}
	    \label{Hamd4parts}
	    &{\mathcal{H}}^{\bf {\rm \bf  h}}_{t}\!= \! \sum\limits_{{\bf k}} \left( {\bf h}_{{\bf k} A}^{\dagger}\hat{V}^{0}_{{\bf k}} {\bf h}_{{\bf k} A}\! +\!{\bf h}_{{\bf k} B}^{\dagger}\hat{V}^{0}_{\bf k} {\bf h}_{{\bf k} B} \right)\! +\! \sum\limits_{{\bf k}, {\bf q}} \left(  {\bf h}_{{\bf k-q} B}^{\dagger} \hat{V}^{\alpha}_{{\bf k},{\bf q}} {\bf h}_{{\bf k} B} \alpha_{\bf q}^{\dagger}  \!+\!
  {\bf h}_{{\bf k-q} A}^{\dagger}  \hat{V}^{\beta}_{{\bf k},{\bf q}} {\bf h}_{{\bf k} B} \beta_{\bf q}^{\dagger} \!+\! h.c. \right)\!,
  \end{align}
  \end{linenomath*}
where (as above) $A, B$ are two AF sublattices, the term $\propto \hat{V}^{0}_{{\bf k}}$ describes the nearest, \kww{next nearest, and third neighbor free hopping, and 
the terms $\propto \hat{V}^{\alpha}_{{\bf k},{\bf q}}$ and  $\propto \hat{V}^{\beta}_{{\bf k},{\bf q}}$ describe the polaronic hopping}.
The detailed derivation and exact expressions for $\hat{V}$'s are again given in SI ~\cite{SM}: while they again depend on the the \kww{five} hopping parameters, we stress that their form is far more complex,
and each $\hat{V}$ is actually a matrix with several nonzero entries.

{\bf Results} Using the SCBA method~\cite{Martinez1991, Liu1992, Sushkov1994, Shibata1999, Wang2015} we 
calculate the relevant Green functions for: (i) the single electron (``$5d^6$ doublon'', $| d \rangle$) doped into the AF ground state of the quasi-2D iridate:
$ G_{\rm IPES} ({\bf k},\omega)=  {\langle {\rm AF} | { d }_{ {\bf k}}\frac{1}{\omega-\mathcal{H}_{\rm IPES}+i\delta} {d}_{ {\bf k}}^\dagger|  {\rm AF} \rangle}$,
and (ii) the single hole (``$5d^4$ hole'', $| {\bf h } \rangle$) doped into the AF ground state of the quasi-2D iridate:
$G_{\rm PES} ({\bf k},\omega)={\rm Tr} {\langle {\rm AF} | { \bf h }_{ {\bf k}}\frac{1}{\omega-\mathcal{H}_{\rm PES}+i\delta}
{\bf h}_{ {\bf k}}^\dagger|  {\rm AF} \rangle}$. We note that using the SCBA method to treat the spin-polaronic problems
is well-established and that the noncrossing approximation is well-justified~\cite{Liu1992, Sushkov1994, Shibata1999}. We solve the SCBA equations on a finite lattice
of $16 \times 16$ sites and calculate the imaginary parts of the above Green's functions -- which (qualitatively) correspond to the theoretical IPES and 
PES spectral functions. 

\begin{figure}[!t]
 \centering
\subfigure{
\includegraphics[width=0.4\linewidth]{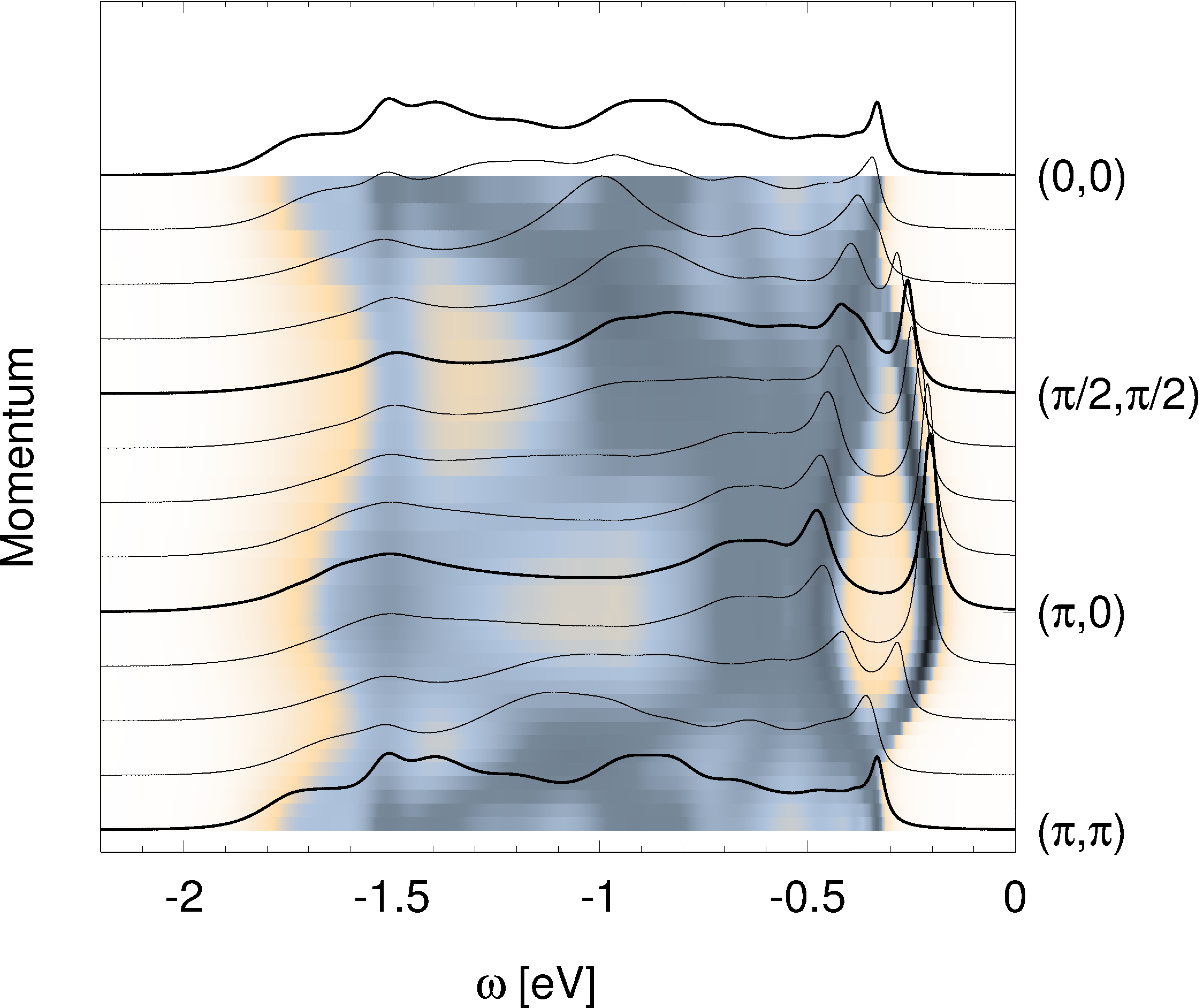}
\llap{
  \parbox[b]{3.85in}{\small{(a) PES}\\\rule{0ex}{2.23in}
  }}\label{fig:2a}%
}
 \subfigure{
\includegraphics[width=0.4\linewidth]{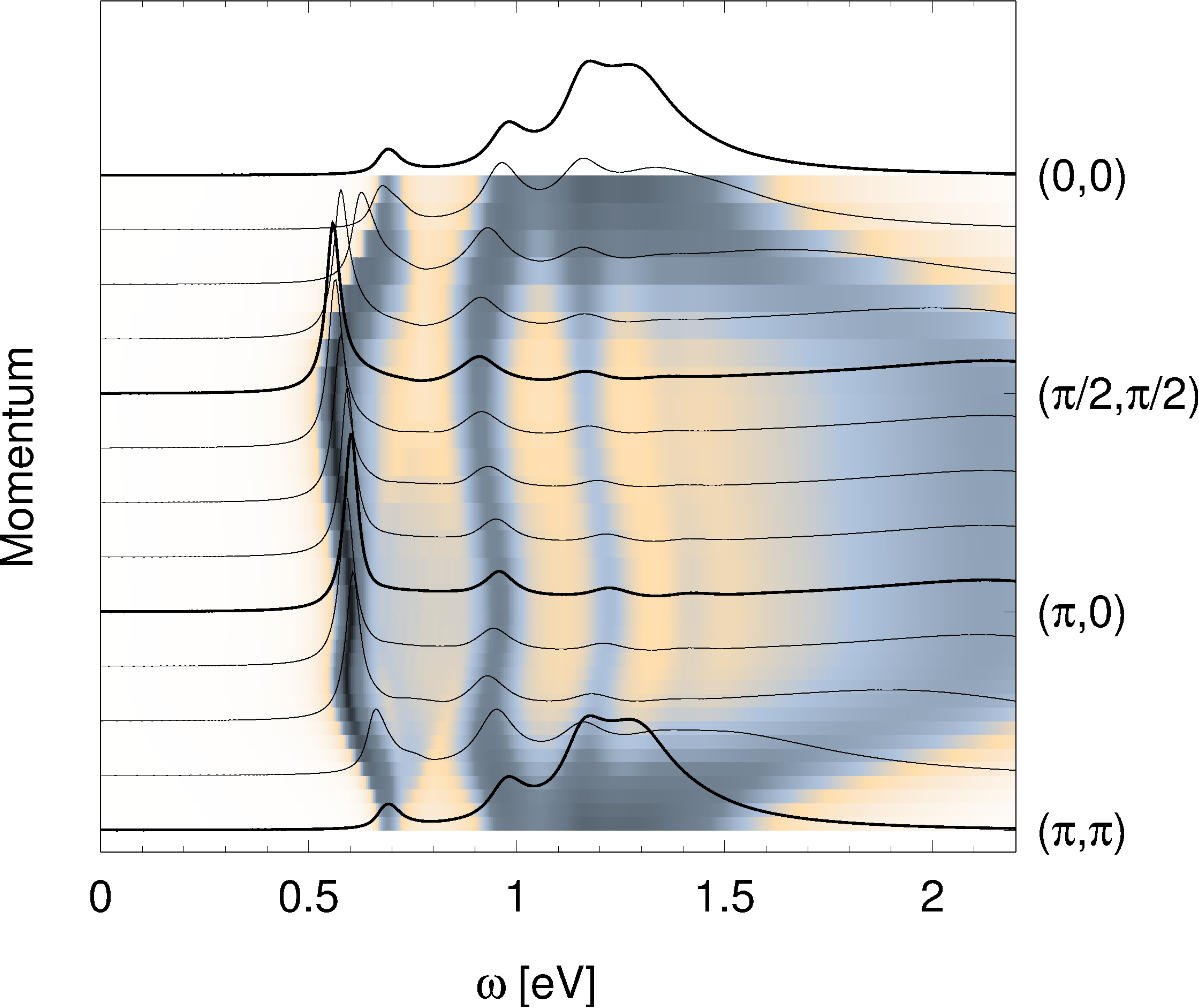}
\llap{
  \parbox[b]{3.8in}{\small{(b) IPES}\\\rule{0ex}{2.23in}
  }}\label{fig:2b}%
}
\caption{
\kww{Theoretical (a) PES and (b) IPES spectral functions for quasi-2D iridates as calculated using SCBA (see text). Parameters: spin exchange $J_1 = 0.06$ eV, $J_2 = - 0.02$ eV, $J_2 = 0.015$ eV,
and spin-orbit coupling $\lambda=0.382$ eV following Ref.~\cite{Naturecom2014Maria}; hopping integrals calculated as the best fit to the DFT data~\cite{SM}
$t_1=-0.2239$ eV, $t_2=-0.373$ eV, $t'=-0.1154$ eV, $t_3=-0.0592$ eV, $t''=-0.0595$ eV; spectra offset by (a) $E=-0.77$ eV and (b) $E=-1.47$ eV. Broadening $\delta = 0.01$ eV.}
\label{fig:2}} 
\end{figure}

We first discuss the calculated angle-resolved IPES spectral function shown in Fig.~\ref{fig:2}(b). One can see that the first addition state has a quasiparticle character, though its dispersion
is relatively small (compared to the LDA bands, see~\cite{SM}): there is a rather shallow minimum
at ($\pi/2$, $\pi/2$) and a maximum at the $\Gamma$ point. Moreover, a large part of the spectral weight is transferred from the quasiparticle
to the higher lying ``ladder'' spectrum, due to the rather small ratio of the spin exchange constants and the electronic hopping~\cite{Martinez1991}. 
Altogether, these are all well-known signatures of the spin-polaron physics: the mobile defect in an AF
is strongly coupled to magnons (leading to the ``ladder'' spectrum) and can move coherently as a quasiparticle only on the scale of the spin exchange $J_1$~\cite{SchmittRink1988, Kane1989, Martinez1991}.
Thus, it is not striking that the calculated IPES spectrum of the iridates is similar to the PES spectrum of the $t$--$J$ model with a ``negative'' next nearest neighbor hopping -- 
the model case of the hole-doped cuprates~\cite{Bala1995, Shibata1999, Damascelli2003, Wang2015}. This agrees with a more general conjecture, previously reported in the literature:
the correspondence between the physics of the hole-doped cuprates and the electron-doped iridates~\cite{WangSenthil2011}.

Due to the internal spin and orbital angular momentum degrees of freedom of the $5d^4$ states, the angle-resolved PES spectrum of the iridates [Fig.~\ref{fig:2}(a)] is very different.
The first removal state shows a quasiparticle character with a relatively small dispersion and a minimum is at the ($\pi$, 0) point (so that we obtain an indirect gap for the quasi-2D iridates). 
On a qualitative level this quasiparticle dispersion resembles the situation found in the PES spectrum of the $t$--$J$ model with a ``positive'' next nearest neighbor hopping~\cite{Wang2015}, which should model the electron-doped cuprates (or IPES on the undoped). However, the higher energy part of the PES spectrum of the iridates is quite distinct not only w.r.t. the IPES but also the PES spectrum of the $t$--$J$ model with the ``positive'' next nearest neighbor hopping~\cite{Bala1995, Damascelli2003, Wang2015}. Thus, the spin-polaron physics, as we know it from the cuprate studies~\cite{SchmittRink1988, Kane1989, Martinez1991}, is modified in this case and we find only very partial agreement with the ``paradigm'' stating that the electron-doped cuprates and the hole-doped iridates show similar physics~\cite{WangSenthil2011}.

\begin{figure}[!t]
 \centering
\subfigure{
\includegraphics[width=0.4\linewidth]{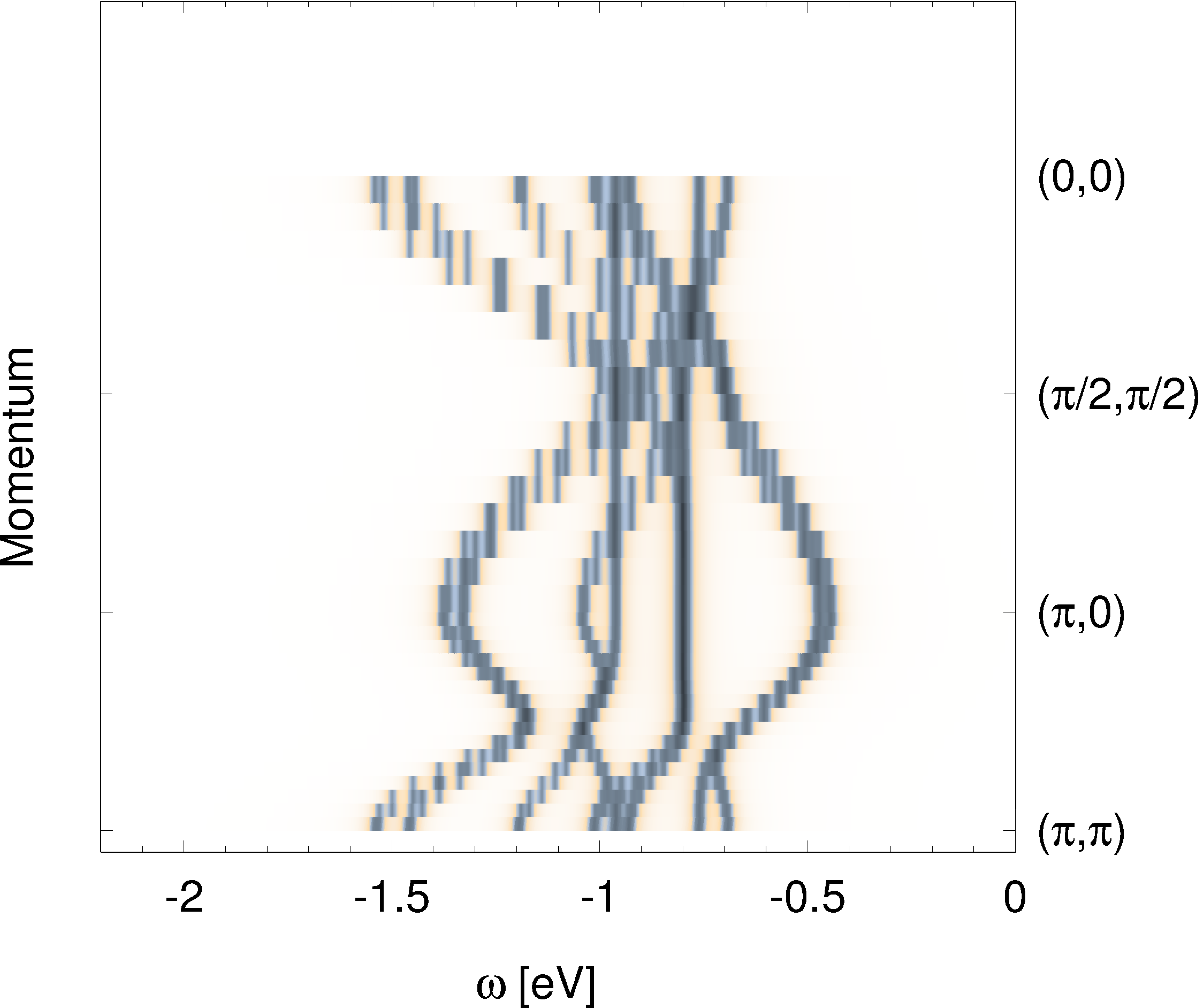}
\llap{
  \parbox[b]{2.9in}{\small{(a) PES:  {free dispersion only} }\\\rule{0ex}{2.23in}
  }}\label{fig:3a}%
}
\subfigure{
\includegraphics[width=0.4\linewidth]{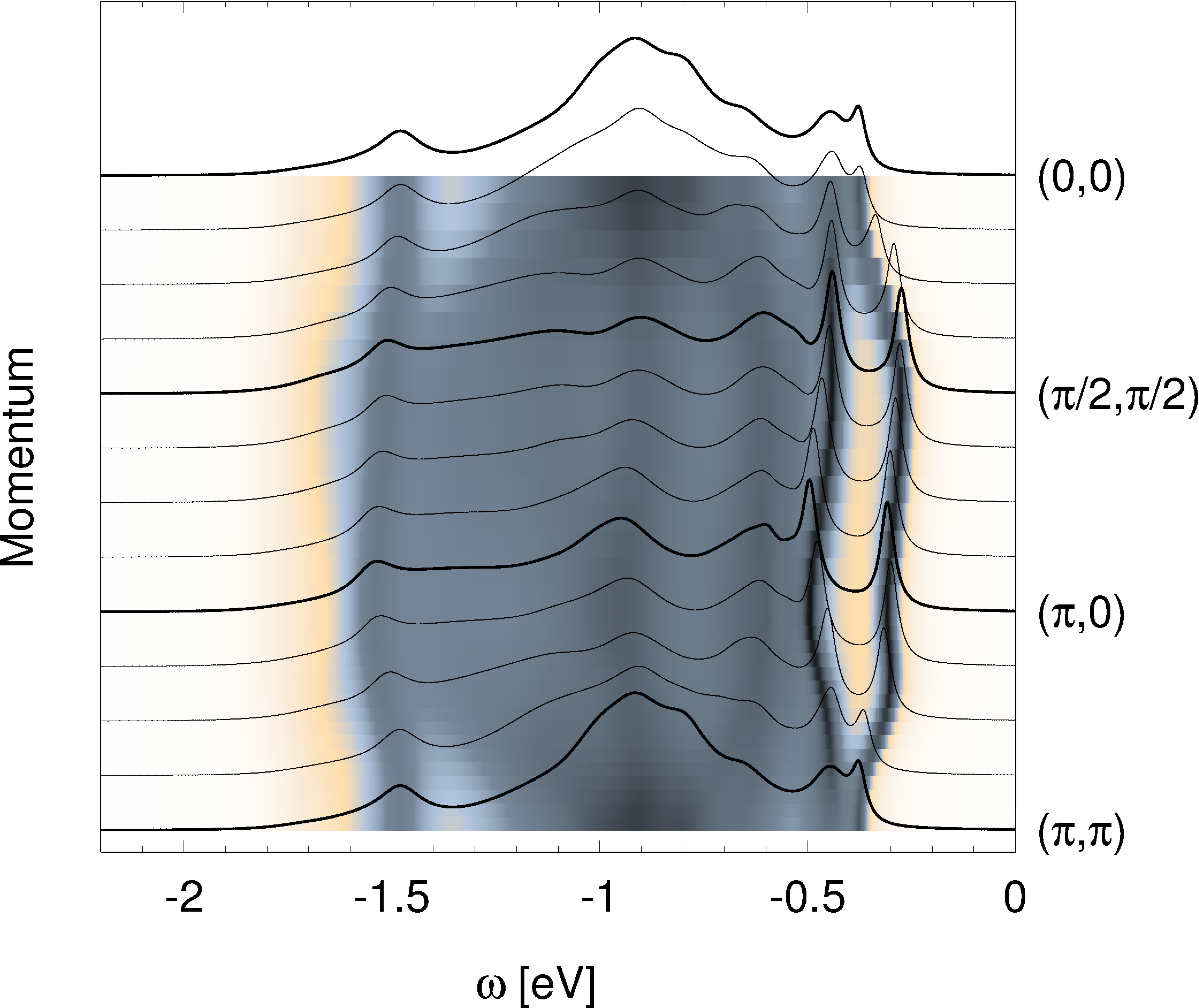}
\llap{
  \parbox[b]{2.9in}{\small{(b) PES: {no free dispersion} }\\\rule{0ex}{2.23in}
  }}\label{fig:3b}%
}
\caption{
Theoretical PES spectral function for quasi-2D iridates (as calculated using SCBA, see text): 
      (a) propagation of the hole not coupled to magnons, i.e. setting $ \hat{V}^{\alpha}_{{\bf k}}=\hat{V}^{\beta}_{{\bf k}}\equiv0$ and
      (b) only polaronic propagation via coupling to magnons (i.e. no free dispersion), i.e. setting $\hat{V}^{0}_{{\bf k}} \equiv 0$.
     Other parameters as in Fig.~\ref{fig:2}.
\label{fig:3}} 
\end{figure}

The above result follows from the interplay between the free [Fig.~\ref{fig:3a}] and polaronic hoppings [Fig.~\ref{fig:3b}] (we note that typically such interplay is highly nontrivial and the resulting full spectrum is never a simple superposition of these two types of hopping processes, cf. Refs.~\cite{Bala1995, Daghofer2008, Wohlfeld2008, Berciu2014, Wang2015, Plotnikova2016}). The free hopping of the ``$5d^4$ hole'' is possible here for both the $J=0$ singlet
and $J=1$ triplets which leads to the onset of several bands. As already stated, the $J=1$ triplets can freely hop not only to the next nearest neighbors but also to the nearest neighbors (see above). For the polaronic hopping, the appearance of {\it several} polaronic channels, originating in the free $J$-bands being dressed by the $j=1/2$ magnons, contributes to the strong quantitative differences w.r.t. the ``$5d^6$ doublon'' case or the cuprates.

\begin{figure}[!t]
 \centering
\subfigure{
\includegraphics[width=0.3\linewidth]{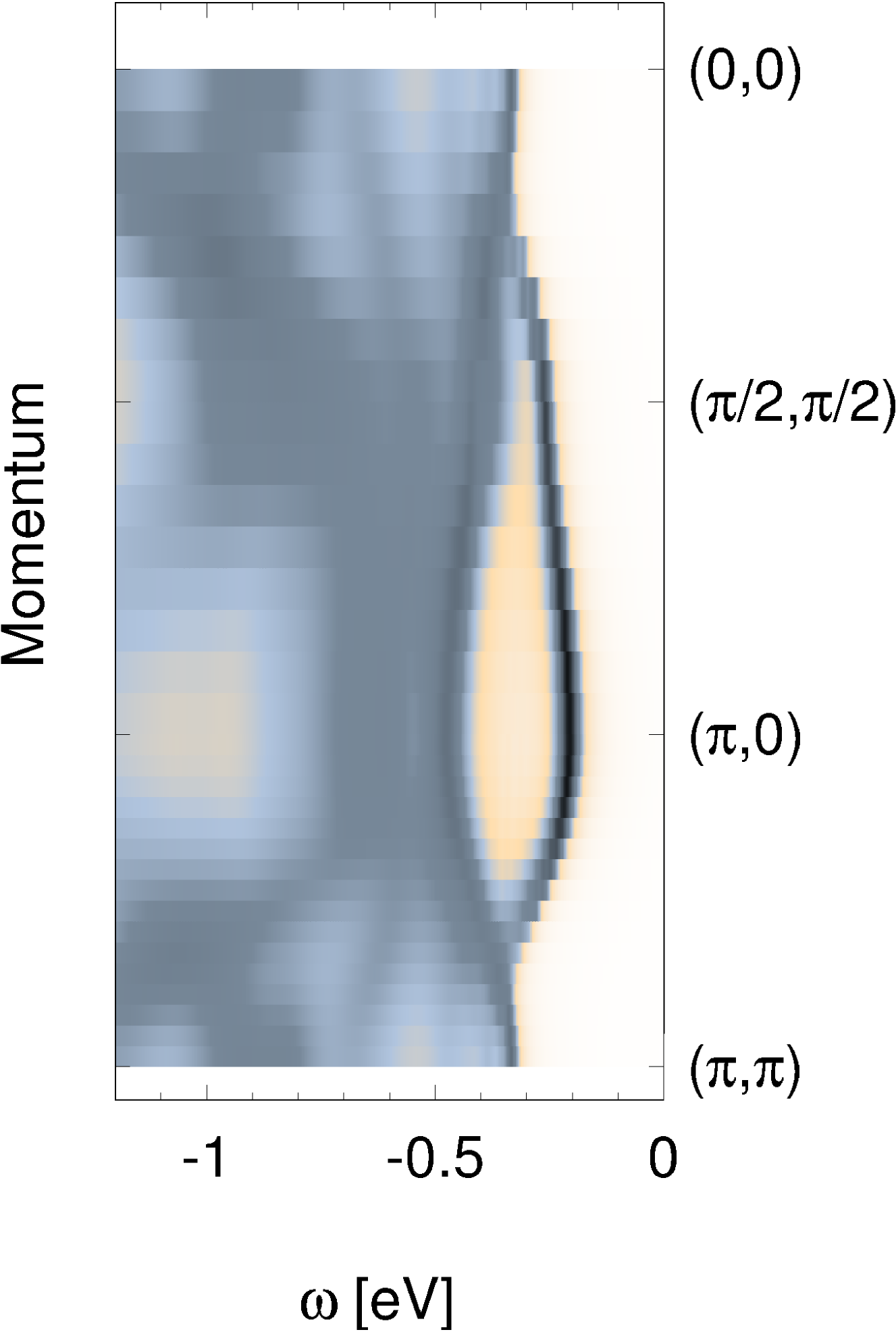}
\llap{
  \parbox[b]{2.8in}{\small{(a)}\\\rule{0ex}{2.9in}
  }}\label{fig:4a}%
}
\subfigure{
\includegraphics[width=0.3\linewidth]{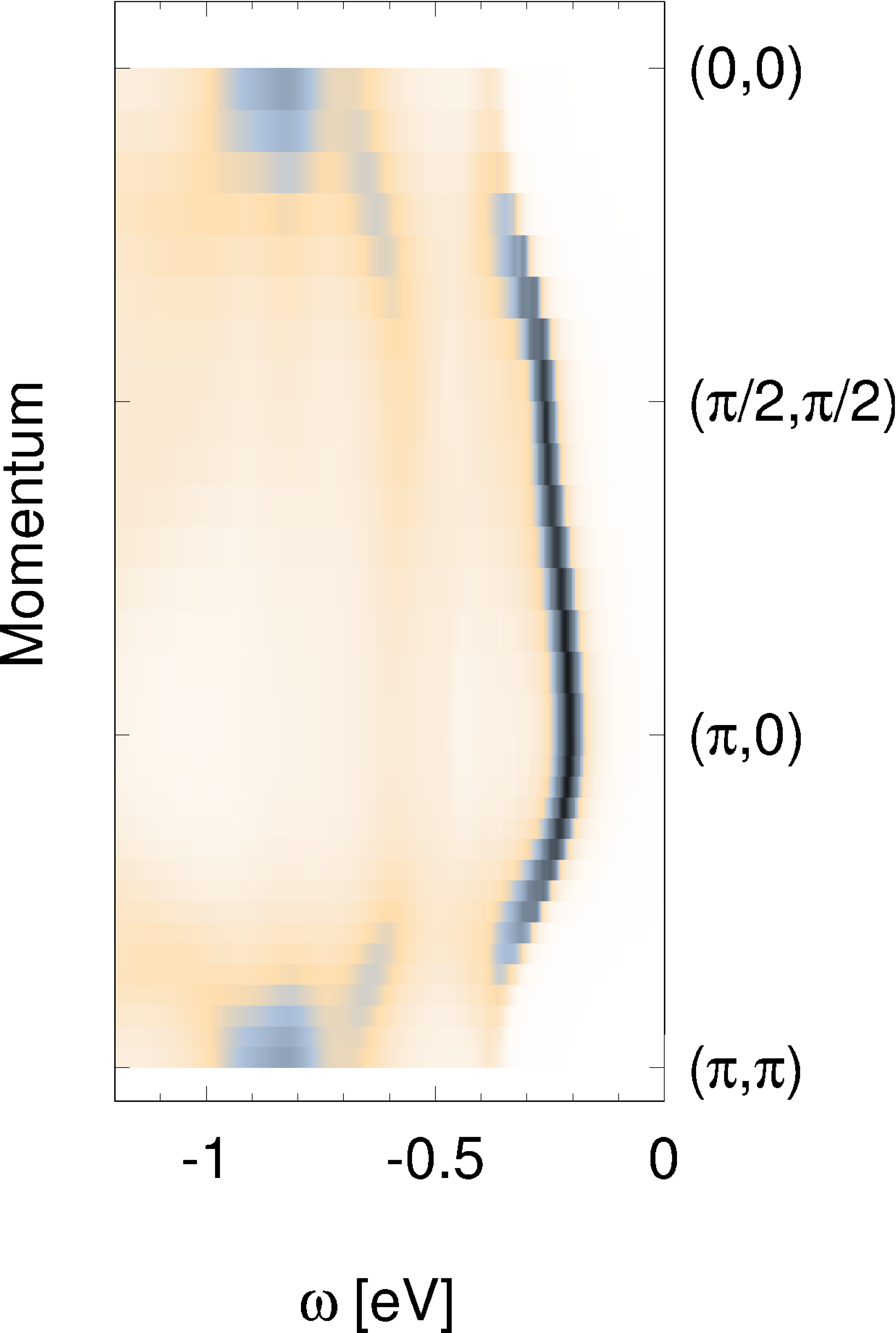}
\llap{
  \parbox[b]{2.5in}{\small{(b) J=0}\\\rule{0ex}{2.9in}
  }}\label{fig:4b}%
}
\subfigure{
\includegraphics[width=0.3\linewidth]{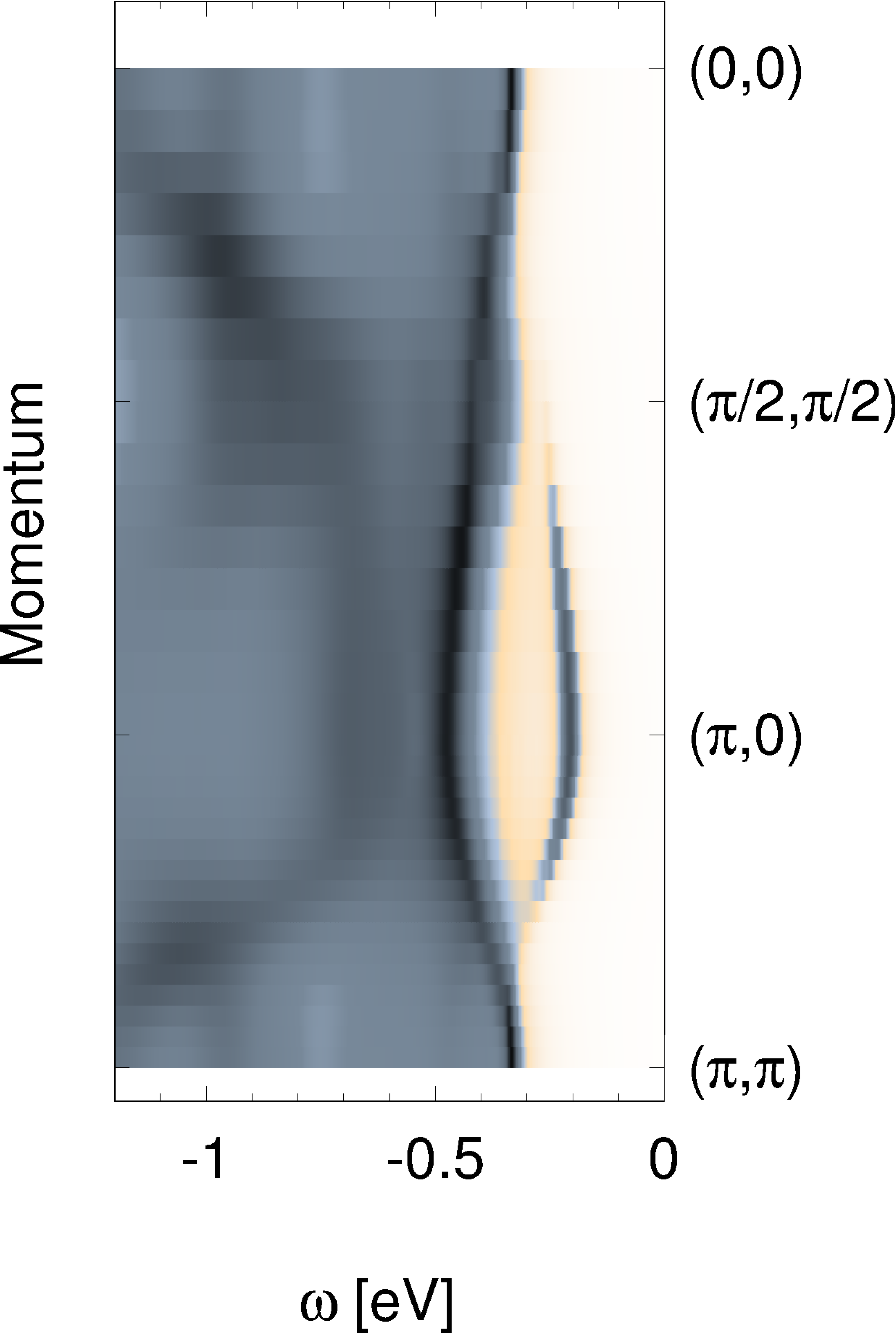}
\llap{
  \parbox[b]{2.5in}{\small{(c) J=1}\\\rule{0ex}{2.9in}
  }}\label{fig:4c}%
}
\caption{A zoom-in into the quasiparticle part of the theoretical PES spectral function for quasi-2D iridates (as calculated using SCBA, see text): the full spectrum (a)
and the $J$-resolved spectra with (b) showing the motion of a ``singlet hole'' (i.e. a hole with $J=0$) and (c) a ``triplet hole'' ($J=1$).
\label{fig:4}} 
\end{figure}

{\bf Comparison with experiment} 
\kww{To directly compare our results with the experimental ARPES spectra of Sr$_2$IrO$_4$~\cite{Kim2008, Nie2015, delaTorre2015, KimNature2016},
we plot the zoomed in spectra for PES, see Fig.~\ref{fig:4}.} 
Clearly, we find the first electron removal state is at a deep minimum at ($\pi$, 0), in good agreement with experiment. 
This locus coincides with the $k$-point where the final state $J=0$  singlet has maximum spectral weight, see Fig.~\ref{fig:4b}. Also the plateau around ($\pi/2$, $\pi/2$) and the shallow minimum of the dispersion at the $\Gamma$ point are reproduced, 
where the latter is related \kp{to} a strong back-bending of higher energy $J=1$ triplets, see Fig.~\ref{fig:4c}. 
Thus one observes that the motion of the ``$5d^4$  hole'' with the singlet character is mostly visible around the minimum at ($\pi$, 0) and near the plateau at ($\pi/2$, $\pi/2$)
[Fig.~\ref{fig:4b}], whereas the triplet is mostly visible at the $\Gamma$ points and much less at ($\pi$, 0) [Fig.~\ref{fig:4c}]. 
The higher energy features in the PES spectrum are mostly of triplet character, due to the difference in the on-site energies between the singlet and triplets $\propto {\lambda}$.
\kww{These features, however, may in case of real materials be strongly affected by the onset of the oxygen states in the PES spectrum (not included in this study, see above).}

Experimentally, electron doping causes Fermi-arcs to appear in Sr$_2$IrO$_4$ that are centered around ($\pi/2$, $\pi/2$)~\cite{Kim2014, KimNature2016, Cao2016, Yan2015,KimNature2016},
which indeed corresponds the momentum at which our calculations place the lowest energy $d^6$ electron addition state. On the basis of the calculated electron-hole asymmetry one expects 
that for hole-doping such Fermi arcs must instead be centered around ($\pi$, 0), unless of course such doping disrupts the underlying host electronic structure of Sr$_2$IrO$_4$.

\kww{Finally, we note that, although the iridate spectral function calculated using LDA+DMFT is also in good agreement with the experimental ARPES spectrum~\cite{Zhang2013}, 
there are two well-visible spectral features that are observed experimentally, and seem to be better reproduced by the current study: 
(i) the experimentally observed maximum at $\Gamma$ point in ARPES being 150-250 meV lower than the maximum at the $X$ point~\cite{Kim2008, delaTorre2015, KimNature2016,  Liu2015, Nie2015,
Brouet2015, Cao2016, Yamasaki2016}, and (ii) the more incoherent spectral weight just below the quasiparticle peak around the $\Gamma$ point than around the ${\rm M}$ point.  We believe that
the better agreement with the experiment 
of the spin polaronic approach than of the DMFT is due to {\it inter alia} the momentum independence of the DMFT self-energy -- which means that the latter method is not able to fully
capture the spin polaron physics~\cite{Martinez1991, Strack1992}.}

{\bf Conclusions}
The differences between the motion of the added hole and electron in the quasi-2D iridates have crucial consequences for
our understanding of these compounds.
The PES spectrum of the undoped quasi-2D iridates should be interpreted
as showing the $J=0$ and $J=1$ bands dressed by $j=1/2$ magnons and a \kww{``free''} nearest and \kww{further} neighbor dispersion. 
The IPES spectrum consists solely of a $J=0$ band dressed by $j=1/2$ magnons and \kww{a ``free'' next nearest and third neighbor dispersion}.
Thus, whereas the IPES spectrum of the quasi-2D iridates qualitatively resemble the PES spectrum of the cuprates, this is not the case of the iridate PES.

This result suggests that, unlike in the case of the cuprates, the differences between the electron and hole doped quasi-2D iridates cannot be modelled
by a mere change of sign in the next nearest hopping in the respective Hubbard or $t$--$J$ model. 
Any realistic model of the hole doped iridates should instead include the onset of $J=0$ and $J=1$ quasiparticle states upon hole doping.

\section{Methods}
The results presented in this work were obtained in two steps: 

Firstly, the proper polaronic Hamiltonians, Eqs. (\ref{Hamd6}) and (\ref{Hamd4}), were derived from the DFT calculations and assuming strong on-site spin-orbit coupling and Coulomb repulsion. This was an analytic work which is described in detailed in~\cite{SM} and which mostly amounts to: (i) the downfolding of the DFT bands to the tight-binding (TB) model, (ii) the addition of the strong on-site spin-orbit coupling and Coulomb repulsion terms to the TB Hamiltonian, and (iii) the implementation of the successive: slave-fermion, Holstein-Primakoff, Fourier, and Bogoliubov transformations. 

Secondly, we calulated the respective Green's functions (see main text for details) for the polaronic model using the self-consistent Born approximation (SCBA). The SCBA is a well-established quasi-analytical method which, in the language of Feynman diagrams, can be understood as a summation of all so-called ``noncrossing'' Feynman diagrams of the polaronic model. It turns out that for
the spin polaronic models (as e.g. the ones discussed here) this approximate method works very well: the contribution of the diagrams with crossed bosonic propagators to the electronic Green's function can be easily neglected~\cite{Liu1992, Sushkov1994, Shibata1999}. Although the SCBA method is in principle an analytical method, the resulting ``SCBA equations'' have to be solved numerically, in order to obtain results which can be compared with the experiment (such as e.g. the spectral functions). The latter was done on a $16 \times 16$ square lattice (the finite size effects are negligible for a lattice of this size).

\section{Acknowledgments}
We would like to thank Krzysztof Byczuk, Ji\v{r}i Chaloupka, Dmitry Efremov, Marco Grioni, Andrzej M Ole\'s, Matthias Vojta and Rajyavardhan Ray for stimulating discussions. K. W. acknowledges support by Narodowe Centrum Nauki (NCN, National Science Center) under Project No. 2012/04/A/ST3/00331. This work has been supported by the Deutsche Forschungsgemeinschaft via SFB 1143.

\section{Author contributions}
E. P. derived the model and performed the SCBA calculations. K. F. performed the LDA calculations. J. v. d. B. and K. W. were responsible for project planning.
K. W., E. P. and J. v. d. B. wrote the paper. 

\section{Author information}
The authors declare no competing financial interests. Correspondence and requests should be addressed to  E. P. (e.plotnikova@ifw-dresden.de).

\newpage

\section{Supplementary Information}
\subsection{A: Magnon dispersion -- detailed form of $\mathcal{H}_{\rm mag}$}
The interaction between the $j=1/2$ isospins in the quasi-2D iridates is well described by the Heisenberg Hamiltonian~\cite{Kim2012}.
Using the successive Holstein-Primakoff, Fourier, Bogoliubov transformations and skipping the terms describing the (iso)magnon interactions, 
we obtain the Heisenberg Hamiltonian in the usual  ``linear spin-wave approximation'' form~\cite{Naturecom2014Maria}:
\begin{linenomath*} \begin{align}
{\mathcal{H}}_{\rm mag}
=\sum\limits_{{\bf q}} \omega_{\bf k} (\alpha^\dag_{\bf q} \alpha_{\bf q} + \beta^\dag_{\bf q} \beta_{\bf k}),
 \end{align} \end{linenomath*}
where $| \alpha_{\bf q} \rangle$ and $ | \beta_{\bf q} \rangle$ are the single magnon states, ${\bf q}$ is the crystal momentum 
and $\omega_{\bf q}$ is the magnon dispersion relation given by
\begin{linenomath*} \begin{align}
\omega_{{\bf q}}=\sqrt{A_{{\bf q}}^2-B_{{\bf q}}^2}
 \end{align} \end{linenomath*}
with
\begin{linenomath*} \begin{align}
A_{\bf q} &=2(J_1-J_2+J_2 \cos q_x \cos q_y-J_3 (1-\frac{1}{2}(\cos 2 q_x +\cos 2 q_y))), \\
B_{\bf q} &=J_1(\cos q_x + \cos q_y).
 \end{align} \end{linenomath*}
Here $J_1$, $J_2$ and $J_3$ are the nearest, next nearest and third neighbor isospin exchange interactions, respectively.

We note at this point that the the parameters of the Bogoliubov transformation, the so-called Bogoliubov coefficients $u_{{\bf q}}$, $v_{{\bf q}}$, are 
given by the following, well-known, expressions in the linear spin-wave theory:
\begin{equation}
\begin{split}
      \label{Bogcoef3}
      & u_{{\bf q}} = \frac{1}{\sqrt{2}}\sqrt{\frac{A_{{\bf q}}}{\omega_{{\bf q}}}+1},\\
      & v_{{\bf q}} = -\frac{sign(B_{{\bf q}})}{\sqrt{2}}\sqrt{\frac{A_{{\bf q}}}{\omega_{{\bf q}}}-1}
\end{split}
\end{equation}
where the coefficients $A_{\bf q}$ and $B_{\bf q}$ are defined above.

\subsection{B: Determining the tight-binding Hamiltonian from the DFT calculations}

\begin{figure}[!h]
\centering
\includegraphics[width=0.25\linewidth]{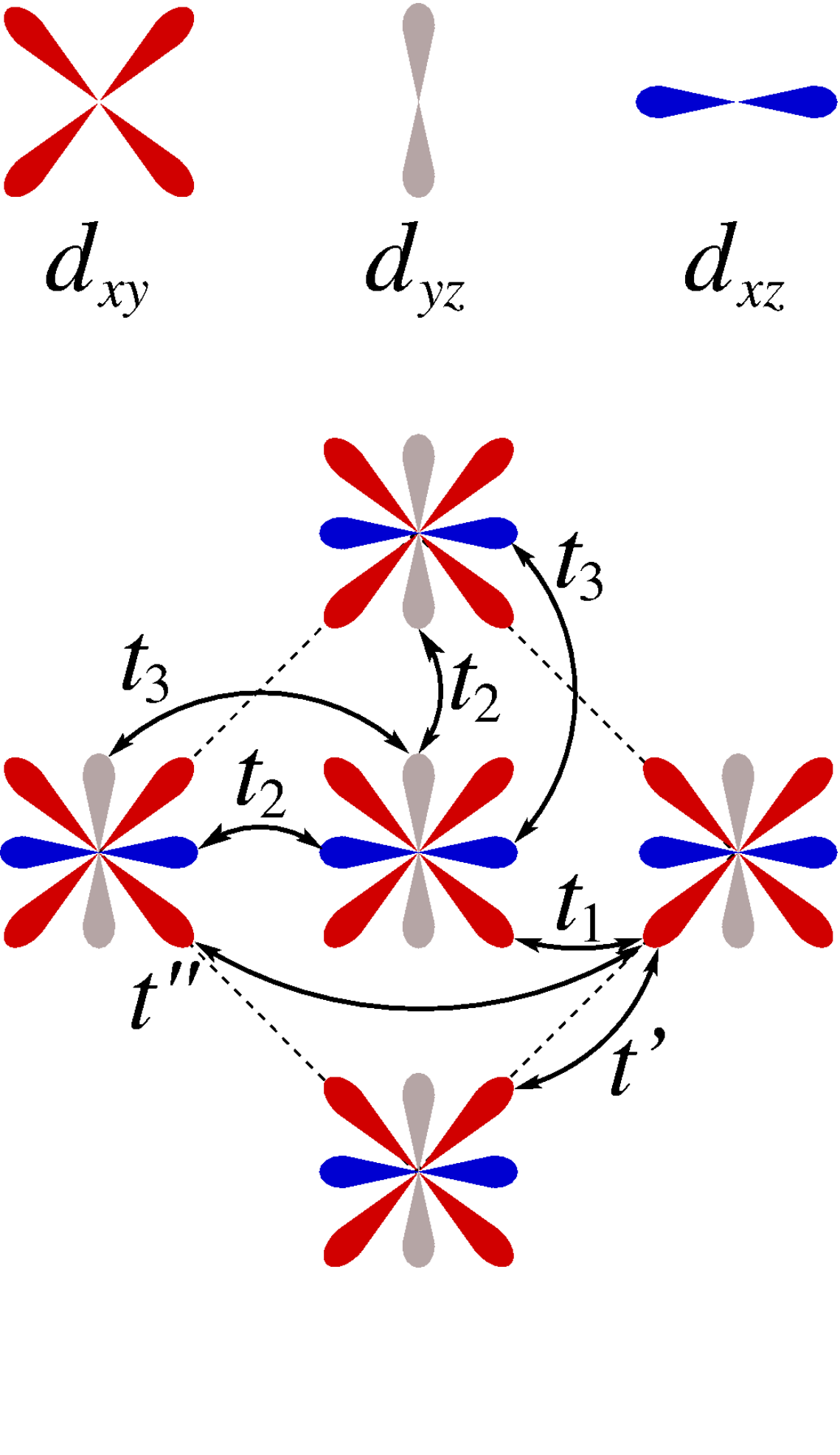}\label{2a}
\llap{
  \parbox[b]{4.3in}{(a)\\\rule{0ex}{2.9in}
  }}
\hspace{1cm}
\includegraphics[width=0.5\linewidth]{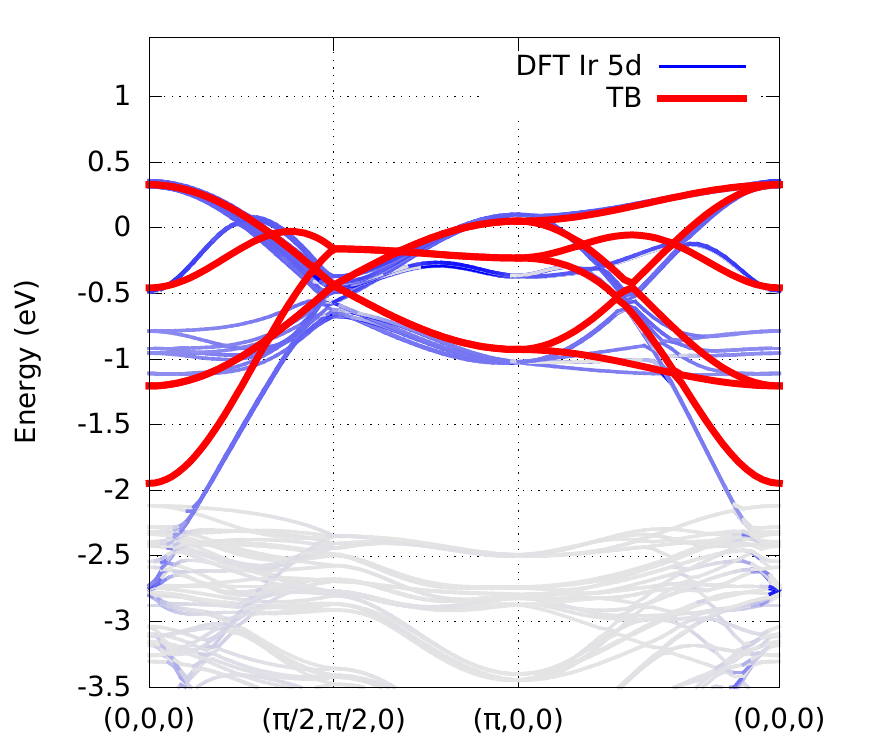}\label{2b}
\llap{
  \parbox[b]{6.5in}{(b)\\\rule{0ex}{2.9in}
  }}
\caption{(a) A cartoon
illustrating the $t_1$, $t_2$, $t_3$, $t'$ and $t''$ hopping
paths between the Ir-$5d$-$t_{2g}$ orbitals.
(b) Comparison
between the DFT (blue) and TB model (red)
dispersion of the Ir-$5d$-$t_{2g}$
bands in Sr$_2$IrO$_4$. The Fermi energy is
set to zero.
\label{2}}
\end{figure}
The electronic band-structure of Sr$_2$IrO$_4$
was calculated using DFT in the local density approximation~\cite{Perdew92}
and within the linearized augmented plane wave approach
using the WIEN2k code~\cite{wien2k}.
We considered the 10~K crystal structure of Sr$_2$IrO$_4$,
with the space group $I 41 {/} acd$, as reported in
Ref.~\cite{Huang1994}.
The calculated band-structure of Sr$_2$IrO$_4$
is shown in Fig.~\ref{2}~(a) along a path in the
Brillouin zone of the $I 41 {/} acd$ unit cell.
The bands with the predominant Ir-$5d$-$t_{2g}$ character
are highlighted in blue.

We used the calculated dispersion of the Ir-$5d$-$t_{2g}$
bands to parameterize our tight-binding (TB) model:
      \begin{linenomath*} \begin{flalign} 
	\label{H_d6NN}
	    \mathcal{H}_{\rm TB}=&
	    -t_1\sum\limits_{\langle {\bf i}, {\bf j} \rangle || {\hat{x}, \hat{y}},\sigma} {c^{\dagger}_{{\bf i} \sigma}c_{ {\bf j} \sigma}}
 	    -t_2\sum\limits_{\langle {\bf i}, {\bf j} \rangle || \hat{y},\sigma} {a^{\dagger}_{{\bf i} \sigma}a_{{\bf j} \sigma}}  
	        -t_2\sum\limits_{\langle {\bf i}, {\bf j} \rangle || \hat{x}, \sigma} {b^{\dagger}_{{\bf i} \sigma}b_{{\bf j} \sigma}}
 	    -t_3\sum\limits_{\langle {\bf i}, {\bf j} \rangle || \hat{y},\sigma} {b^{\dagger}_{{\bf i} \sigma}b_{{\bf j} \sigma}}  \nonumber \\
	        &-t_3\sum\limits_{\langle {\bf i}, {\bf j} \rangle || \hat{x}, \sigma} {a^{\dagger}_{{\bf i} \sigma}a_{{\bf j} \sigma}}
	        	    -t' \sum\limits_{\langle \langle {\bf i}, {\bf j} \rangle \rangle || \hat{x'}, \hat{y'} ,\sigma} {c^{\dagger}_{ {\bf i} \sigma}c_{ {\bf j} \sigma}}
	        	   -t'' \sum\limits_{\langle \langle \langle {\bf i}, {\bf j} \rangle \rangle \rangle || \hat{x''}, \hat{y''} ,\sigma} {c^{\dagger}_{{\bf i} \sigma}c_{ {\bf j} \sigma}} + h.c.,
  \end{flalign} \end{linenomath*}
where $a^{\dagger}$, $b^{\dagger}$, and $c^{\dagger}$ operator create an electron in the $d_{yz}$, $d_{xz}$, $d_{xy}$ orbitals (respectively) with spin $\sigma=\pm \frac{1}{2}$, 
$\hat{x}$ and $\hat{y}$ indicate the directions of the nearest neighbor bonds in the $xy$ plane of the quasi-2D iridate, and $\hat{x'}= \hat{x}-\hat{y}$ and $\hat{y'}= \hat{x}+\hat{y}$ 
($\hat{x''}= 2\hat{x}$ and $\hat{y'}= 2\hat{y}$)
indicate the directions of the next nearest (third) neighbor bonds in the $xy$ plane of the quasi-2D iridate.
The TB model includes the nearest neighbor
hopping integrals $t_1$, $t_2$ and $t_3$ as well as the 
next nearest and third neighbor integral $t'$ and $t''$ between the
Ir-$5d$-$t_{2g}$ orbitals, with their meaning explained in Fig.~\ref{2}~(a).

We found that the following parameter values are both physically
reasonable and give a satisfactory match
between the DFT and TB model bands:$t_1=-0.2239$ eV, $t_2=-0.373$ eV, $t'=-0.1154$ eV, $t_3=-0.0592$ eV,$t''=-0.0595$ eV.
The TB model band-structure based
on these parameter values is shown
in red in Fig.~\ref{2}~(b).
Let us also note that the generic structure of the TB Hamiltonian follows from the well-known symmetries of an effective TB Hamiltonian 
for the transition metal oxide with the $t_{2g}$ orbital degrees of freedom: the electrons located in the $d_{ab}$ orbital can solely hop in the $ab$ plane.

\subsection{C: Motion of the ``$5d^6$ doublon'' -- detailed form of  ${\mathcal{H}}_{t}^{ {\rm d}}$}

\kww{Having obtained the TB Hamiltonian we are now ready to derive the Hamiltonian which would describe the motion of the  ``$5d^6$ doublon'' added to the Mott insulating ground state formed
by the $5d^5$ iridium ions of the (undoped) quasi-2D iridates due to the nonzero hopping elements of the TB Hamiltonian. This means that the main task here is to calculate the following matrix  
elements of the tight-binding Hamiltonian [Eq. (6) above] $\langle 5d^6_{\bf i} 5d^5_{\bf j} |  \mathcal{H}_{\rm TB}  | 5d^5_{\bf i} 5d^6_{\bf j}\rangle$. This is done in several steps:}

\kww{Firstly, we calculate the above matrix elements in the appropriate eigenstates of ionic Hamiltonian of the $5d^5$ and $5d^6$ configurations (these states are listed in Fig. 1. of the main text). 
We note that these matrix elements do not explicitly depend on the strong on-site spin-orbit coupling $\lambda$, though the form of the appropriate eigenstates of the ionic Hamiltonian (Fig. 1 of the main text) is of course
due to the onset of strong on-site spin-orbit coupling $\lambda$. Secondly, we assume the so-called no double occupancy constraint, which follows from the implicitly assumed here
limit of strong on-site Coulomb repulsion -- which prohibits the creation of ``unnecessary''  ``$5d^6$ doublons'' once the electron added to the quasi-2D iridate $5d^5$ ground state hops between
sites. Technically this amounts to the introduction of the projection operator which takes care of this constraint.
Finally, following the path described for example in Refs.~\cite{Martinez1991,Plotnikova2016} and introducing the slave-fermion formalism followed by 
Fourier and Bogoliubov transformations, we arrive at the following polaronic Hamiltonian which describes the motion of the ``$5d^6$ doublon'':}
  \begin{linenomath*} \begin{align}
	    \label{Hamd6SI}
	    &{\mathcal{H}}_{t}^{\bf {\rm d}} = \sum\limits_{{\bf k}}{V^{0}_{{\bf k}} \left(d^{\dagger}_{{\bf k} A}d_{{\bf k} A} + d^{\dagger}_{{\bf k} B}d_{{\bf k} B}\right)}+ \sum\limits_{{\bf k},{\bf q}} V_{{\bf k},{\bf q}}
	    \left(d^{\dagger}_{{\bf k-q} B}d_{{\bf k} A} \alpha_{\bf q}^{\dagger}+d^{\dagger}_{{\bf k-q} A}d_{{\bf k} B} \beta_{\bf q}^{\dagger} +h.c. \right),
 \end{align} \end{linenomath*}
with the free next-nearest and third- neighbor hopping
    \begin{linenomath*} \begin{flalign} 
   &V^0_{\bf k}=-\frac{4t'}{3}\gamma'_{{\bf k}}-\frac{4t''}{3}\gamma''_{{\bf k}},
    \end{flalign} \end{linenomath*}
and the vertex
    \begin{linenomath*} \begin{flalign} 
   &V_{{\bf k},{\bf q}}=-\frac{8(t_1+t_2+t_3)}{3\sqrt{2N}}\left(\gamma_{{\bf k-q}}u_{\bf q}+\gamma_{{\bf k}}v_{\bf q}\right),
    \end{flalign} \end{linenomath*}
   where $\gamma_{{\bf k}}=1/2(\cos{k_x}+\cos{k_y})$, $\gamma'_{{\bf k}}=\cos{k_x}\cos{k_y}$  and $\gamma''_{{\bf k}}=1/2(\cos{2 k_x}+\cos{2 k_y})$ and the Bogoliubov coefficients $u_{\bf q}$ and $v_{\bf q}$ are given in Sec.~A above.
\subsection{D: Motion of the ``$5d^4$ hole'' -- detailed form of  ${\mathcal{H}}_{t}^{\bf  {\rm h}}$}

The Hamiltonian which describes the motion of the ``$5d^4$ hole'' ( ${\mathcal{H}}_{t}^{\bf  {\rm h}}$) is derived in a similar way as in the  ``$5d^6$ doublon''  case described in Sec.~C.
However, due to the \kww{multiplet structure of the eigenstates of the ionic Hamiltonian of the} $5d^4$ configuration (see Fig. 1 of the main text), its form is far more complex -- in the low energy limit it describes the hopping
of the four distinct eigenstates that can be formed by the ``$5d^4$ hole''  (singlet $S$, and three triplets $T_\sigma$; see main  text):
\begin{linenomath*} \begin{align}
	    \label{Hparts}
	    &{\mathcal{H}}^{\bf {\rm \bf  h}}_{t}\!= \! \sum\limits_{{\bf k}} \left( {\bf h}_{{\bf k} A}^{\dagger}\hat{V}^{0}_{{\bf k}} {\bf h}_{{\bf k} A}\! +\!{\bf h}_{{\bf k} B}^{\dagger}\hat{V}^{0}_{\bf k} {\bf h}_{{\bf k} B} \right)\! +\! \sum\limits_{{\bf k}, {\bf q}} \left(  {\bf h}_{{\bf k-q} B}^{\dagger} \hat{V}^{\alpha}_{{\bf k},{\bf q}} {\bf h}_{{\bf k} B} \alpha_{\bf q}^{\dagger}  \!+\!
  {\bf h}_{{\bf k-q} A}^{\dagger}  \hat{V}^{\beta}_{{\bf k},{\bf q}} {\bf h}_{{\bf k} B} \beta_{\bf q}^{\dagger} \!+\! h.c. \right)\!,
   \end{align} \end{linenomath*}
with the free hopping
\begin{linenomath*} \begin{align}
     \label{V0}
      &\hat{V}^{0}_{\bf k}= \begin{pmatrix}
      F_1 & 0 & -F_2 & 0 &   0 & P_2 & 0 & -P_1\\
      0 & F_4 & 0 & 0 &      P_1 & 0 & Q_1 & 0 \\
      -F_2 & 0 & F_3 & 0 &   0 & Q_2 & 0 & Q_1 \\
      0 & 0 & 0 & 0 &        -P_2 & 0 & Q_2 & 0 \\
      0 & P_1 & 0 & -P_2 &   F_1 & 0 & F_2 & 0\\
      P_2 & 0 & Q_2 & 0  &   0 & 0 & 0 & 0  \\
      0 & Q_1 & 0 & Q_2 &    F_2 & 0 & F_3 & 0   \\
      -P_1 & 0 & Q_1 & 0 &   0 & 0 & 0 & F_4  \\
      \end{pmatrix},
 \end{align} \end{linenomath*}
and the vertices
\begin{linenomath*} \begin{align}
      &\hat{V}^{\alpha}_{{\bf k},{\bf q}}= \begin{pmatrix}
      0 & L_3 & 0 & -L_3 &  Y_1 & 0 & -W_2 & 0\\
      L_3 & 0 & L_1 & 0 &  0 & Y_4 & 0 & W_1 \\
      0 & L_1 & 0 & L_1 &  -W_2 & 0 & Y_2 & 0 \\
      -L_3 & 0 & L_1 & 0 &  0 & W_1 & 0 & Y_3 \\
      0 & 0 & 0 & 0   & 0 & L_4 & 0 & -L_4 \\
      0 & 0 & 0 & 0   & L_4 & 0 & L_2 & 0  \\
      0 & 0 & 0 & 0   & 0 & L_2 & 0 & L_2 \\
      0 & 0 & 0 & 0   & -L_4 & 0 & L_2 & 0  \\
      \end{pmatrix}, \nonumber
 \end{align} \end{linenomath*}
\begin{linenomath*} \begin{align}
      &\hat{V}^{\beta}_{{\bf k},{\bf q}}= \begin{pmatrix}
      0 & L_4 & 0 & -L_4 &  0 & 0 & 0 & 0 \\
      L_4 & 0 & L_2 & 0 &  0 & 0 & 0 & 0 \\
      0 & L_2 & 0 & L_2 &  0 & 0 & 0 & 0 \\
      -L_4 & 0 & L_2 & 0 &  0 & 0 & 0 & 0 \\
     Y_1 & 0 & W_2 & 0   & 0 & L_3 & 0 & -L_3  \\
     0 & Y_3 & 0 & W_1   & L_3 & 0 & L_1 & 0  \\
     W_2 & 0 & Y_2 & 0   & 0 & L_1 & 0 & L_1  \\
     0 & W_1 & 0 & Y_4   & -L_3 & 0 & L_1 & 0  \\
      \end{pmatrix}.
       \end{align} \end{linenomath*}

The nearest neighbor free hopping $P({\bf k})$, $Q({\bf k})$ and the polaronic diagonal $Y({\bf k},{\bf q})$ and non-diagonal $W({\bf k},{\bf q})$ vertex elements are 
   \begin{linenomath*} \begin{flalign} 
   &P_1({\bf k})=\frac{2\left(2t_1-t_2\right)}{3\sqrt{3}}\gamma_{\bf k}-\frac{2t_3}{3\sqrt{3}}\gamma_{\bf k}, \\
   &P_2({\bf k})=\frac{2t_2}{\sqrt{3}}\tilde{\gamma}_{\bf k}-\frac{2t_3}{\sqrt{3}}\tilde{\gamma}_{\bf k},\\
   &Q_1({\bf k})=\frac{\left(4t_1+t_2\right)}{3\sqrt{2}}\gamma_{\bf k}+\frac{t_3}{3\sqrt{2}}\gamma_{\bf k},\\
      &Q_2({\bf k})=\frac{t_2}{\sqrt{2}}\tilde{\gamma}_{\bf k}-\frac{t_3}{\sqrt{2}}\tilde{\gamma}_{\bf k},\\
   &W_1({\bf k},{\bf q})=\frac{t_3-t_2}{\sqrt{2N}}\left(\tilde{\gamma}_{\bf k-q}u_{\bf q}+\tilde{\gamma}_{{\bf k}}v_{\bf q}\right),\\
   &W_2({\bf k},{\bf q})=-\frac{4\left(2t_1-t_2-t_3\right)}{3\sqrt{3N}}\left(\gamma_{{\bf k-q}}u_{\bf q}-\gamma_{{\bf k}}v_{\bf q}\right),\\
   \end{flalign} \end{linenomath*}
   \begin{linenomath*} \begin{flalign} 
   &Y_1({\bf k},{\bf q})=-\frac{16\left(t_1+t_2+t_3\right)}{9\sqrt{2N}}\left(\gamma_{{\bf k-q}}u_{\bf q}+\gamma_{{\bf k}}v_{\bf q}\right),\\
   &Y_2({\bf k},{\bf q})=-\frac{2\left(4t_1+t_2+t_3\right)}{3\sqrt{2N}}\left(\gamma_{{\bf k-q}}u_{\bf q}+\gamma_{{\bf k}}v_{\bf q}\right),\\
   &Y_3({\bf k},{\bf q})=-\frac{4t_1+t_2+t_3}{3\sqrt{2N}}\gamma_{{\bf k}}v_{\bf q}-\frac{3\left(t_2+t_3\right)}{\sqrt{2N}}\gamma_{{\bf k-q}}u_{\bf q},\\
   &Y_4({\bf k},{\bf q})=-\frac{4t_1+t_2+t_3}{3\sqrt{2N}}\gamma_{{\bf k-q}}u_{\bf q}-\frac{3\left(t_2+t_3\right)}{\sqrt{2N}}\gamma_{{\bf k}}v_{\bf q},
   \end{flalign} \end{linenomath*}
   where $\tilde{\gamma}_{{\bf k}}=1/2(\cos{k_x}-\cos{k_y})$.
  The free hopping elements arising from the next-nearest and third neighbor hoppings are:
   \begin{linenomath*} \begin{flalign} 
   &F_1({\bf k})=-\frac{4 t'\gamma'_{\bf k}}{9}-\frac{4 t''\gamma''_{\bf k}}{9},\\
   &F_2({\bf k})=-\frac{8 t'\gamma'_{\bf k}}{3\sqrt{6}}-\frac{8 t''\gamma''_{\bf k}}{3\sqrt{6}},\\
   &F_3({\bf k})=-\frac{2t'\gamma'_{\bf k}}{3}-\frac{2t''\gamma''_{\bf k}}{3},\\
   &F_4({\bf k})=-\frac{t'\gamma'_{\bf k}}{3}-\frac{t''\gamma''_{\bf k}}{3}.
    \end{flalign} \end{linenomath*}
    And the polaronic next-nearest and third neighbor hopping elements are:
   \begin{linenomath*} \begin{flalign} 
   &L_1({\bf k},{\bf q})=\frac{4 t'}{3\sqrt{N}}\gamma'_{{\bf k-q}}u_{\bf q}+\frac{4 t''}{3\sqrt{N}}\gamma''_{{\bf k-q}}u_{\bf q},\\
   &L_2({\bf k},{\bf q})=\frac{4 t'}{3\sqrt{N}}\gamma'_{{\bf k}}v_{\bf q}+\frac{4 t''}{3\sqrt{N}}\gamma''_{{\bf k}}v_{\bf q},\\
   &L_3({\bf k},{\bf q})=\frac{8 t'}{3\sqrt{6N}}\gamma'_{{\bf k-q}}u_{\bf q}+\frac{8 t''}{3\sqrt{6N}}\gamma''_{{\bf k-q}}u_{\bf q},\\
   &L_4({\bf k},{\bf q})=\frac{8 t'}{3\sqrt{6N}}\gamma'_{{\bf k}}v_{\bf q}+\frac{8 t''}{3\sqrt{6N}}\gamma''_{{\bf k}}v_{\bf q}.
    \end{flalign} \end{linenomath*}

\clearpage
\bibliography{arpes_sr2iro4_v4}
\end{document}